\documentclass[letterpaper,prl,twocolumn,showpacs]{revtex4}
\usepackage{physics}
\usepackage{amsmath,amssymb}
\usepackage{epsfig}
\usepackage[geometry]{ifsym}
\usepackage{color}
\usepackage{fixltx2e}

\newcommand {\ra}{\rightarrow}

\newcommand {\pt}{\partial}
\newcommand{\I}{\mathrm i}

\begin{document}
\title{New approaches to coding information using inverse scattering transform}
\date{\today}
\author{L.\,L.~Frumin $^{1,2}$}
\author{A.\,A.~Gelash$^{2,3}$}
\author{S.\,K.~Turitsyn$^{2,4}$}\email{s.k.turitsyn@aston.ac.uk}
\affiliation{$^{1}$Institute of Automation and Electrometry SB RAS, Novosibirsk, Russia}
\affiliation{$^{2}$Novosibirsk State University, Novosibirsk, Russia}
\affiliation{$^{3}$Kutateladze Institute of Thermophysics, SB RAS, Novosibirsk, Russia}
\affiliation{$^{4}$Aston Institute of Photonics Technologies, Aston, Birmingham, United Kingdom}
\begin{abstract}
Remarkable mathematical properties of the integrable nonlinear Schr\"{o}dinger equation (NLSE)
can offer advanced solutions for the mitigation of nonlinear signal distortions in optical fibre links.
Fundamental optical soliton, continuous and discrete eigenvalues of the nonlinear spectrum have already been
considered for transmission of information in fibre-optic channels.
Here we propose to apply signal modulation to the kernel of the Gelfand-Levitan-Marchenko
equations (GLME) that offers the advantage of a relatively simple decoder design.
First, we describe an approach based on exploiting the general N-soliton solution of the NLSE for simultaneous coding of $N$ symbols
involving $4\times N$ coding parameters. As a specific elegant sub-class of the general schemes we introduce a soliton
orthogonal frequency division multiplexing (SOFDM) method. This method is based on the choice of identical imaginary
parts of the N-soliton solution eigenvalues, corresponding to equidistant soliton frequencies making it similar to the conventional
OFDM scheme, thus, allowing use of the efficient fast Fourier transform algorithm to recover the data.
Then, we demonstrate how to use this new approach to control signal parameters in the case of the continuous spectrum.
\end{abstract}
\maketitle
\section{Introduction}
The nonlinear Schr\"{o}dinger equation (NLSE) is a generic
fundamental mathematical model with numerous applications in
science and technology. In particular, the NLSE describes a
path-average propagation of light in fibre-optic systems that is
the backbone of the modern global communication networks.
The NLSE is an example of a fundamental
model of nonlinear physics which can be integrated by the Inverse
Scattering Transform (IST) method~\cite{zakharov1972exact,AKNS}.
The IST method is one of the greatest achievements of
mathematical physics in the 20-th century (see e.g.~\cite{zakharov1972exact,AKNS,ZMNP,Newell,AS,hasegawa1995solitons,lamb1980elements}
and references therein). In recent years (especially in optical communications) the
IST method is also referred to as the Nonlinear Fourier Transform (NFT) stressing the similarity to the
conventional Fourier transform and the ability of the
IST/NFT to present solutions of the nonlinear evolution equation
on the basis of non-interacting modes called scattering data
or (in the NFT notation) \textit{nonlinear spectrum}.

One specific, albeit highly important, application of the NLSE is in optical communications, where
it is derived as a path-average (over periodic variations of power due to loss and gain) model governing
the signal propagation along the transmission line~\cite{hasegawa1995solitons,kivshar2003optical,mollenauer2006solitons} (here we use the normalised version):
\begin{equation}\label{NLSE}
\I\frac{\pt q}{\pt z} + \frac{1}{2}\frac{\pt^2 q}{\pt t^2} +  |q|^2q=0.
\end{equation}
Here $q(z,t)$ is an optical field envelope, $z$ is a propagation
distance along the optical fibre and $t$ is the retarded time.

The general solution of the NLSE is presented by the superposition of solitary (localised in time) waves
corresponding to the discrete (solitonic) part of the nonlinear spectrum and dispersive waves associated with the
continuous part of the nonlinear spectrum. Recent advances in coherent optical communication allowing information coding both over amplitude and phase
have made it possible to reconsider relatively old ideas of using the soliton solution of the NLSE~\cite{hasegawa1995solitons,mollenauer2006solitons} and nonlinear spectrum eigenvalues for the transmission of information~\cite{HN} in the new context. The recent surge of interest in nonlinear transmission techniques is in particular due to the observation that conventional (linear) data transmission techniques are facing serious challenges induced by the nonlinear properties of the optical fibre communication channels (an excellent overview is given in~\cite{Rene1,Rene2}). This calls for the development of new nonlinear techniques of signal coding, transmission and processing.

The traditional soliton transmission has been recently reassessed in~\cite{PDTPRL,yushko2014coherent} in the context of coherent communication and the use of soliton phase for data transmission. Moreover, a great deal of interest has been sparked recently by the application of the powerful IST/NFT methods in optical communications (see e.g.~\cite{NFT1,yousefi2014information,NFT3,NFT4,NFT5,NFT6,NFT7} and references therein, we simply are not able to review here all relevant papers that have been published recently in this fast growing field).

The efficiency of numerical algorithms for data encoding/decoding is critically important
in the digital telecommunication networks. For instance, in wireless communication, the success and popularity of the orthogonal frequency division multiplexing
(OFDM) method is due to the exceptional computational performance and high spectral efficiency
of the fast Fourier transform (FFT)~\cite{shieh2009ofdm}.
The success of the practical implementation of the nonlinear IST/NFT techniques will be defined by the availability of the fast and super-fast NFT methods~\cite{belai2007efficient,frumin2015efficient,NFT5} and the stability of algorithms with respect to noise impact. The IST/NFT technique is relatively new compared to conventional methods and the currently available numerical algorithms of information encoding/decoding using solitonic signal are still far from the efficiency required in practical hardware implementation.

Here we propose to use the kernel of the Gelfand-Levitan-Marchenko equations (GLME) to encode information, in particular, we demonstrate that the OFDM scheme can be applied in an efficient way. To create signal at the beginning of the transmission line as well as to recover the encoded kernel at the end of the line, here we use the efficient Toeplitz inner-bordering (TIB) numerical scheme of inverse and direct scattering transform (which was recently introduced by Frumin and co-authors~\cite{belai2007efficient,frumin2015efficient}) and the exact soliton solution known from the IST theory \cite{zakharov1972exact,AKNS}.

For the discrete nonlinear spectrum, we propose a soliton orthogonal frequency division multiplexing (SOFDM) technique that is
based on the choice of identical imaginary parts of N-soliton solution eigenvalues, corresponding to equidistant soliton frequencies making it similar to the conventional OFDM scheme and allowing the use of the efficient fast Fourier transform algorithm to recover the data.
We also demonstrate how the concept of the OFDM can be applied for the continuous spectrum kernel~\cite{NFT3}. The important advantage of using coding over kernel of the GLME is the possibility of controlling signal parameters by utilising the time domain window functions in the modulated kernel.

 \section{$N$-soliton solutions of the NLSE for $N$-symbol block transmission}
In the traditional soliton transmission a single (soliton) pulse is used as an information carrier
sent over a time slot allocated to one symbol in a given spectral channel~\cite{hasegawa1995solitons,mollenauer2006solitons}. Transmission in this case is affected by the soliton interactions, and/or is restricted in the spectral efficiency, because a separate soliton occupies a small fraction of the symbol duration time.
A great deal of attention has recently been placed on the potential use of the discrete nonlinear eigenvalues in fibre-optic channels.
 Most of the current studies of discrete nonlinear eigenvalue communications are limited to exploring different solitonic waveforms (forming a transmitted alphabet) in a single symbol time slot. To avoid interaction between neighbouring symbols a long guard interval is typically used to suppress inter-symbol interactions, thus, limiting spectral efficiency of such burst-mode transmission.

We propose here to use the well-known~\cite{zakharov1972exact,AKNS}
general analytical $N$-soliton solutions of the NLSE (N-SS) (see the Supplemental Material~\cite{SupM}) for $N$-symbol block modulation and coding. In a block transmission technique the information symbols are arranged in the blocks separated by some known symbols. Application of the N-SS allows one to simultaneously code information over $N$ symbol time intervals. Four soliton parameters in principle offer a possibility of four-dimensional modulation/coding per soliton/symbol.
Over the interval of $N$ symbols, $N$-soliton solutions can offer $4 \times N$ degrees of freedom.

Recall that single soliton solutions reads:
\begin{equation}\label{1-SS}
q^{(1)}(z,t) = 2\beta \frac{\exp[-2i\omega t - 2i (\omega^2-\beta^2) z + i \theta]}{\cosh(2\beta t + 4\omega\beta z - \delta t)}\,.
\end{equation}
Here, obviously, $2 \beta$ corresponds to soliton amplitude, $2 \omega$ is soliton frequency, $\theta$ is pulse phase
and $ \delta t/(2 \beta)$ defines soliton timing position. These four parameters can be used for coding of information, i.e. amplitude, frequency, phase and pulse position modulations, leading to various high-level modulation formats. Note that interactions between solitons are automatically accounted for in the $N$-soliton solution. Therefore, in N-SS coding there is no issue of soliton interactions that occur when solitons are treated as separate entities.

The N-SS is defined by its scattering data/nonlinear spectrum: two sets of $N$ complex constants. The first set corresponds to the complex eigenvalues of
solitons $\xi_k = i\beta_k+\omega_k$, $k=1,...,N$. As discussed above, the imaginary part $\beta_k>0$ defines corresponding (with the index $k$) soliton amplitude and the real part $\omega_k$ is related to the soliton frequency (and corresponding group speed). The second set is given by the complex numbers $c_k=C_k \exp(i\theta_k)$ with real parameters $C_k$ and $\theta_k$. For the well separated solitons, parameters $C_k$ define timing positions of solitons in the following way: $\delta t_k = \ln[C_k/(2\beta_k)]$, while parameters $\theta_k$ define soliton phases. Based on the structure of the solitonic scattering data, the possible data coding of N-SS form two natural groups classified as: \textit{amplitude-frequency} modulation and \textit{pulse position-phase} modulation. In general, there are $4 \times N$ free parameters that can be used for modulation.

The generation of modulated (i.e. encoded) N-SS signal at the transmitter requires an algorithmic realisation of IST/NFT in the encoder. Here, to find the N-SS we use the standard factorisation of the GLME, that leads to the well known exact formulae (see the Supplemental Material~\cite{SupM}). Alternatively, the N-SS can be obtained by algebraic versions of IST such as the Zakharov-Shabat dressing
method~\cite{zakharov1974scheme}, Darboux transformation~\cite{matveev1991darboux}, method of
Hirota~\cite{hirota1973exact} and by the IST TIB algorithm. Note, that all these approaches are numerically unstable at large $N$, that limits their applications (see the Supplemental Material~\cite{SupM}). The kernel $\Omega(z,t)$ of the GLME for the N-SS has the following form:
\begin{equation}\label{kernel}
\Omega^{(N)}(t,z) = \sum_{k=1}^{N} c_k(z)e^{-i\xi_k t}\,.
\end{equation}
The sum in~Eq.(\ref{kernel}) is similar to the Fourier series,
however, the "frequencies" $\xi_k$, in general, are complex numbers. Formally,
the N-SS can be written as IST/Inverse NFT of the kernel~(\ref{kernel}):
\begin{equation}\label{N-SS IST}
q^{(N)}(z,t) = IST[\Omega^{(N)}(z,t)]\,.
\end{equation}
In what follows, for the sake of simplicity, the index $(N)$ will be
 omitted. We assume that coding/modulation is applied at
$z=0$ (encoder) and decoding/demodulation (decoder) at $z=L$. The
IST/NFT method links the nonlinear spectrum $z=0$ and $z=L$ by the
following simple phase shift:
\begin{equation}\label{phase correction}
c_k(L) = c_k(0)\exp(-2i\xi_k^2 L), \,\,\,\,\, k=(1,...,N)\,.
\end{equation}
Expressions~(\ref{N-SS IST}) and~(\ref{phase correction})
formally solve the problem of compensation of signal distortions in the communication channels
described by the NLSE. We believe, that formula~(\ref{kernel}) can offer some advantages for
encoding/decoding operations compared to the traditionally used formula~(\ref{N-SS IST}).

Our central idea is to use the N-SS
kernel~(\ref{kernel}) for the modulation of the information data. In this case the
number of numerical operations at the decoder is reduced. In the proposed scheme
the decoding operation requires to recover only the kernel~(\ref{kernel}) at $z=0$ by solving the direct scattering
problem and by application of a simple transformation~(\ref{phase
correction}). Moreover, as we will demonstrate, the analogy of the
N-SS kernel~(\ref{kernel}) with the Fourier series allows to
introduce the OFDM scheme for the discrete spectrum.

\section{Solitonic OFDM method}

In this section we introduce a soliton OFDM (SOFDM) technique in which the GLME kernel can be efficiently used for encoding/decoding $2\times N$ position-phase parameters.  The key idea can be understood from the expression for the N-soliton kernel~(\ref{kernel}).
The kernel would be similar to the conventional OFDM in case of real $\xi_k$.
Therefore, we impose special conditions on the complex soliton parameters $\xi_k$. Namely, we consider
N-SS with eigenvalues $\xi_n = \omega_n + iA$. In this case
solitons have equal amplitudes, but different equidistantly selected frequencies. Thus, the
GLME kernel~(\ref{kernel}) at the beginning of the line is given
by the finite Fourier series multiplied by $e^{At}$:
\begin{equation}\label{SOFMkernel}
\Omega(0,t) = e^{At} \sum_{k=1}^{N} c_k e^{-i \omega_k t}\,.
\end{equation}
This greatly simplifies the processing of such signals.

Now, without loss of generality, we consider modulation over phase $\theta_n$,
while the pulse positions $\delta_n$ are left unmodulated. As a particular
example of the SOFDM encoding we consider
$\tilde{N}$-phase-shift keying ($\tilde{N}$-PSK) modulated N-SS.
To apply the SOFDM over the finite time slot $T$ we introduce the
discrete time grid:
\begin{equation}
 t_m = (m-1)T/N, \,\,\, m=1,...,N.
 \end{equation}
The orthogonality of Fourier harmonics is given by the following
condition:
\begin{equation}\label{Orthogonality}
t_m \omega_n = 2\pi (m-1)(n-1)/N, \,\,\, m,n=1,...,N\,.
\end{equation}
Similar to the standard OFDM, the FFT makes it possible to determine
parameters of signal modulation $c_n$ by $O(N \ln N)$ arithmetic
operations:
\begin{equation}
c_n = FFT[\Omega(0,t_m) \exp(-A t_m)]\,.
\end{equation}

To compute the scattering data from the received signal $q(t,L)$ one can use any available algorithm of the direct NFT. Here, without loss of generality, we use the direct TIB algorithm calculating the entire signal kernel in time domain by solving the GLME (see Supplemental Material~\cite{SupM} and ~\cite{frumin2015efficient}). The kernel contains all scattering data information: soliton eigenvalue positions (corresponding to amplitude and frequency modulations), pulse positions and phases ($4\times N$ parameters). Here we focus only on a phase modulation to illustrate the proposed concept. The eigenvalue modulation is also possible, but it faces challenges in terms of efficiency and stability  (see Supplemental Material~\cite{SupM}).

For illustration purposes we choose the minimum possible number of time samples $M=N$. But
actually, the value of $M$ depends on the algorithm of the direct
scattering transform at the receiver and usually $M > N$.
At the transmitter, we use the IFFT to obtain the
kernel~(\ref{SOFMkernel}) from the given data encoded by the phases $c_n$ and
then solve the inverse scattering problem as described in the Supplemental Material~\cite{SupM}
to generate the input optical N-SS signal $q(0,t_m)$.

\begin{figure}[h]
\includegraphics[width=1.65in]{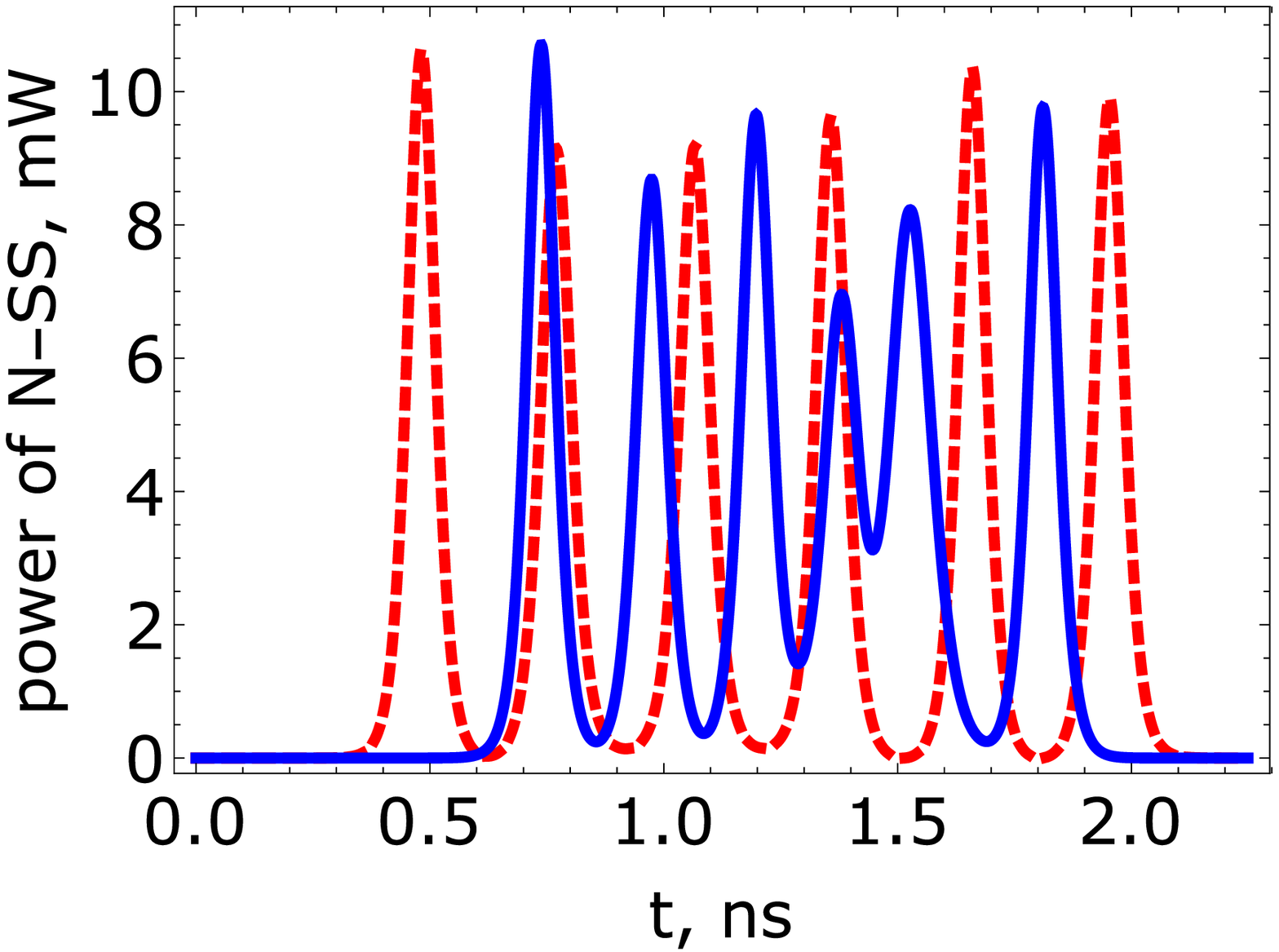}
\includegraphics[width=1.65in]{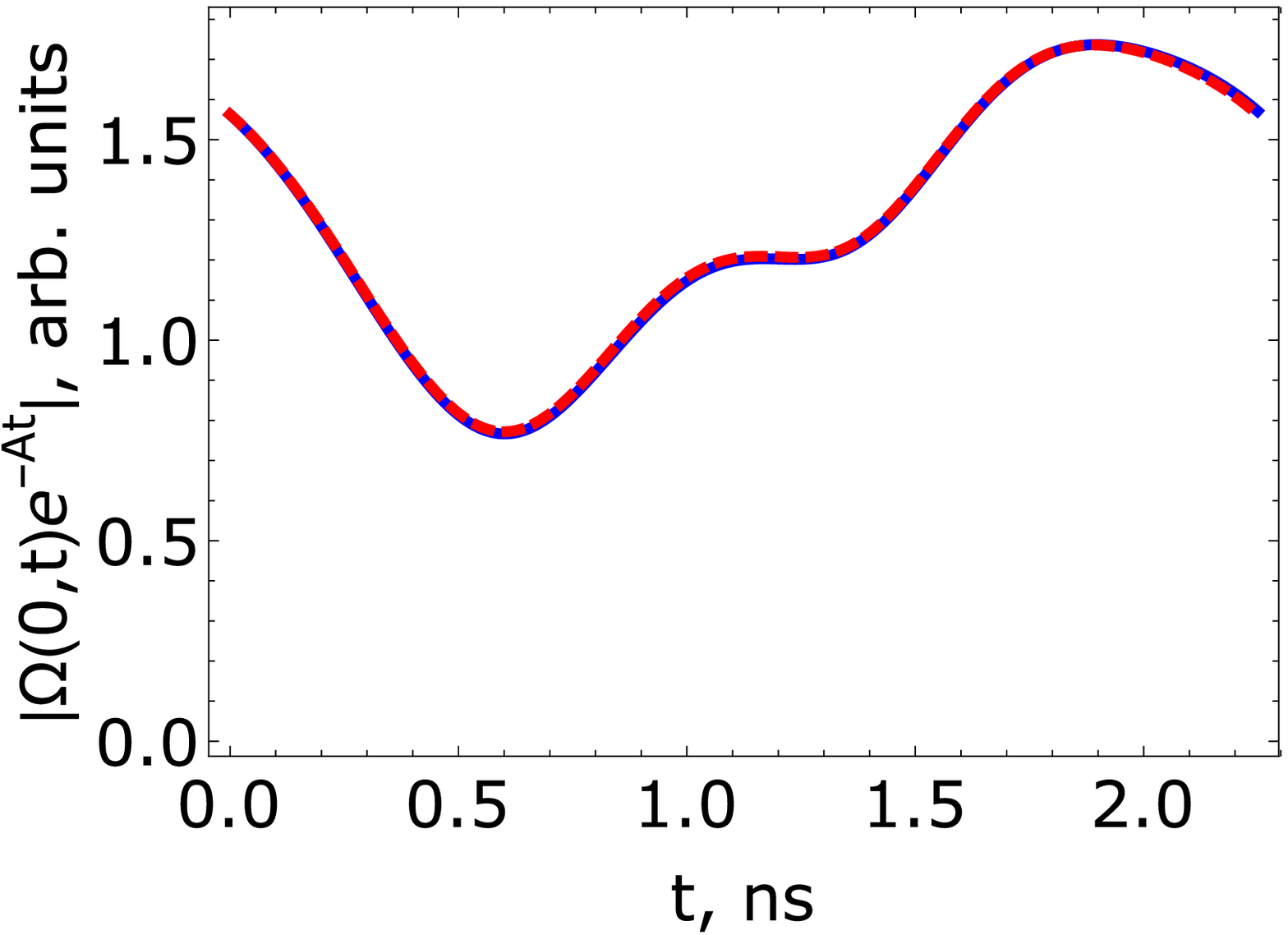}
\caption{\label{Fig_1} LEFT: The modulus of 6-soliton solution at
the beginning ($q_0$, red dashed line) and at the end ($q_1$, blue solid line) of the NLSE governed transmission line.
\\
RIGHT: 6-soliton normalized kernel of the GLME
equations at the beginning of the transmission line:
$|\Omega(0,t)exp(-At)|$. Blue solid line -- the exact normalized
kernel encoded by the SOFDM method with QPSK, red dashed line -- the same
kernel restored by the use of direct TIB method.}
\end{figure}

We test the SOFDM method in numerical simulations of data
transmission by the use of quaternary phase-shift keying (QPSK)
modulated $6$-SS. In Fig.~\ref{Fig_1} (left) we present an example of
a 6-soliton signal at the beginning and at the end of the transmission line of length $L=2000$ km.
Using the direct TIB method we recover the encoded kernel that presented in Fig.~\ref{Fig_1} (right).
To avoid signal expansion, we arrange the solitons in order of descending velocity so that the slowest soliton occupies
the first position in the signal while the fastest soliton starts propagation from the signal end.
 However, we would like to stress that the practical implementation of the solitonic OFDM scheme requires further development of fast noise-stable methods for solving the direct scattering problem that we consider in the Discussion section.

\section{Kernel coding of the continuous nonlinear spectrum OFDM}

The kernel of the GLME for the continuous spectrum is presented in the following form:
\begin{equation}\label{CSOFMkernel}
\Omega^{(R)}(0,t) = R(0,t) = \frac{1}{2\pi}\int_{-\infty}^{\infty}r_{\omega}e^{-i\omega t} d\omega \,.
\end{equation}
Here $r_{\omega}$ is the reflection coefficient of the Zakharov-Shabat system for the given signal $q^{(R)}(0,t)$ while $R(0,t)$ is commonly referred to as a signal response function.  Similar to the discrete spectrum case, the IST links the continuous spectrum at $z=0$ and $z=L$ by the following relation:
\begin{equation}\label{phase correction CS}
r_{\omega}(L) = r_{\omega}(0)\exp(-2i\omega^2 L)\,.
\end{equation}

The general idea to apply the OFDM scheme to the continuous nonlinear spectrum was previously considered in the framework of the so-called "nonlinear inverse synthesis" method~\cite{NFT3,le2015nonlinear}. This approach, whilst promising, has an important challenge -- how to control the signal characteristics in the time domain? Indeed, the reflection coefficient $r(\omega)$ which was chosen for encoding information is nontrivially coupled with the signal via IST.

Here, we propose to apply the additional window transformation to the kernel $\Omega^{(R)}(0,t)$ in the time domain, as a method of controlling signal parameters. For IST-based schemes the strong localisation of signal in time slots is highly critical to avoid nonlinear interactions between neighbouring symbol intervals. Bearing in mind the linear limit ($q^{(R)}(z,t) \ra 2\Omega^{(R)}(z,t)$) we conclude, that well localised (in time domain) signals should correspond to the localised kernel at least in a weakly nonlinear case. We have examined different window transformation functions known from the linear communication theory (see, for instance~\cite{jenkins1968spectral}) and found that the excellent signal localisation in time is achieved for window functions with smooth polynomial fronts.

\begin{figure}[h]
\includegraphics[width=1.65in]{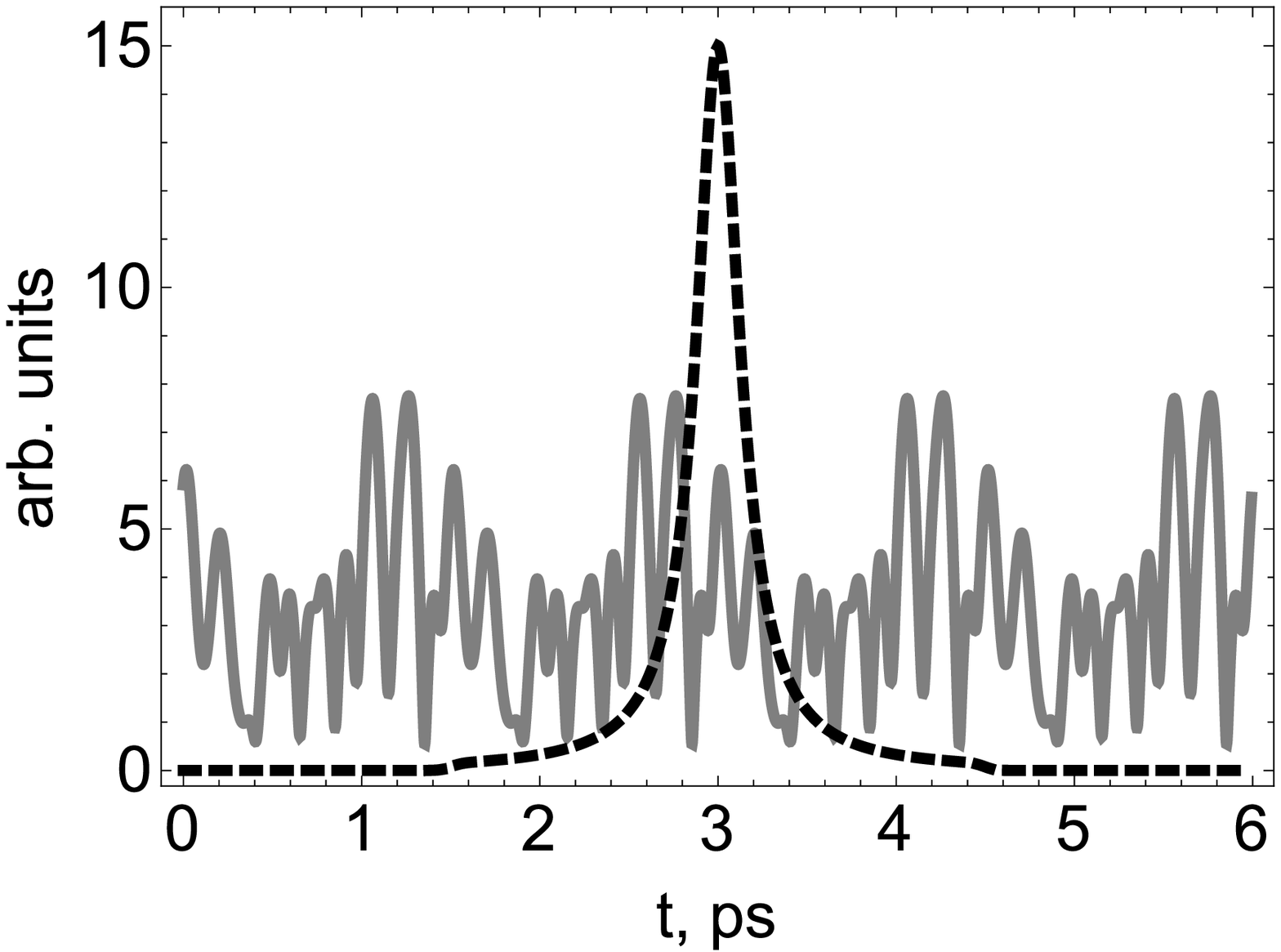}
\includegraphics[width=1.65in]{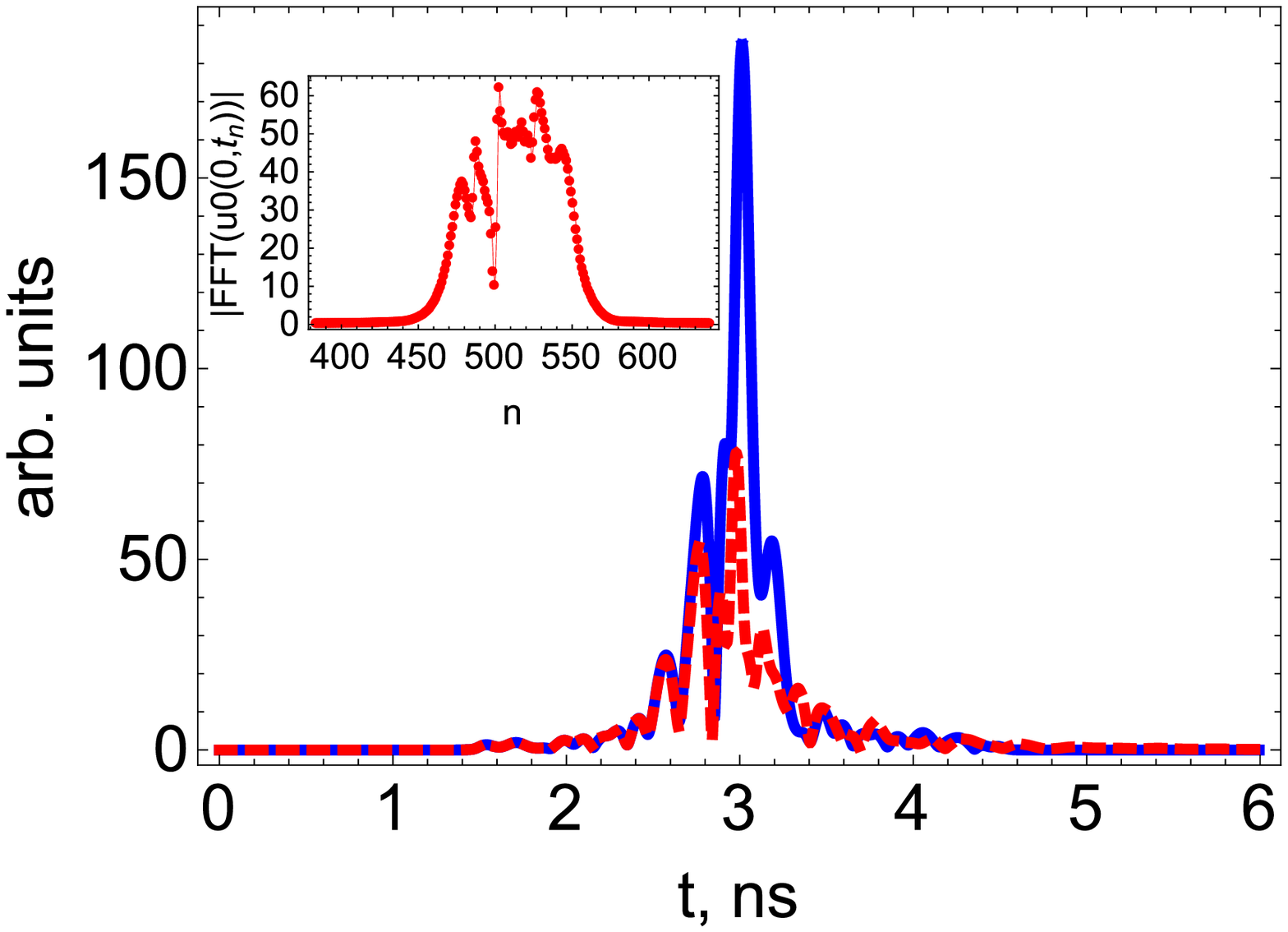}
\caption{\label{Fig_2} LEFT: 16 encoded kernel harmonics (grey, solid line) and window function (black, dashed line). Parameters of the window transformation function~(\ref{Lorenz}) are the following: $\tilde{A}=15,\, \Gamma=20$.
\\
RIGHT: The double kernel (blue, solid line) obtained by multiplying the encoded harmonics by the window function and corresponding signal obtained by the use of the inverse TIB algorithm (red, dashed line). The inset picture shows  the  absolute value of signal Fourier spectrum, frequency index $n$ is defined in ~(\ref{Orthogonality}))}.
\end{figure}

Figure~\ref{Fig_2} demonstrates signal generation at the beginning of the transmission line. We start from 16 Fourier harmonics encoded using the OFDM $8$-PSK scheme. Then we apply the window transformation $f(t)$ similar to the well known Lorentzian function:
\begin{equation}\label{Lorenz}
F(t) = \frac{\tilde{A}}{ [\Gamma(t-t_0)]^2 + 1}\,,
\end{equation}
to localise signal in time slot. Here $\tilde{A},\, \Gamma$ are the parameters of the window transformation corresponding to the characteristic amplitude and width of the modulated kernel and $t_0$ corresponds to the centre of the time slot. We also add to the window transformation function~(\ref{Lorenz}) exponentially decaying tails which do not affect the general shape of the signal, but helps to cancel interactions between neighbouring bursts. Finally, we find the signal profile using the inverse TIB method (Fig.~\ref{Fig_2}, right, red).

\begin{figure}[h]
\includegraphics[width=1.65in]{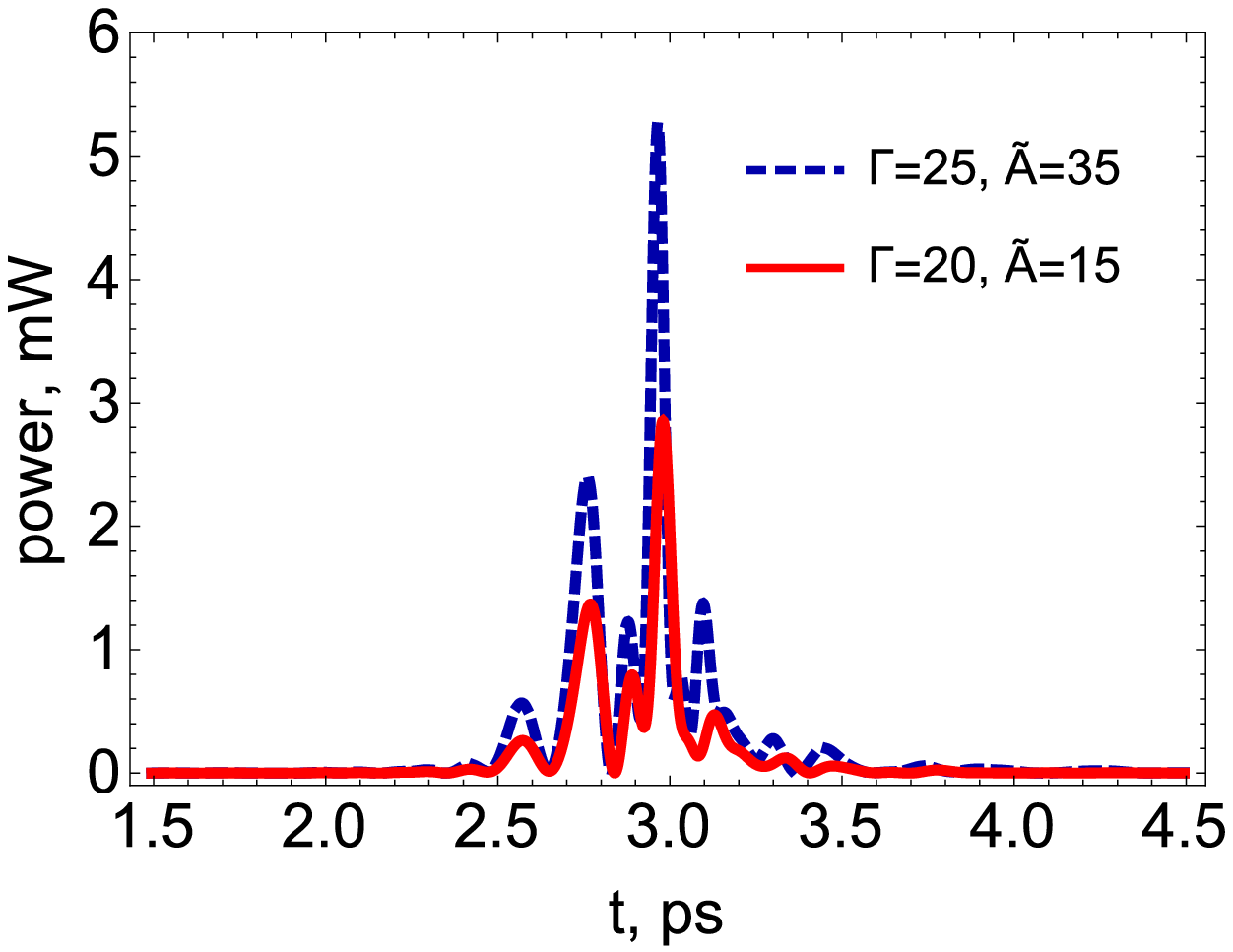}
\includegraphics[width=1.65in]{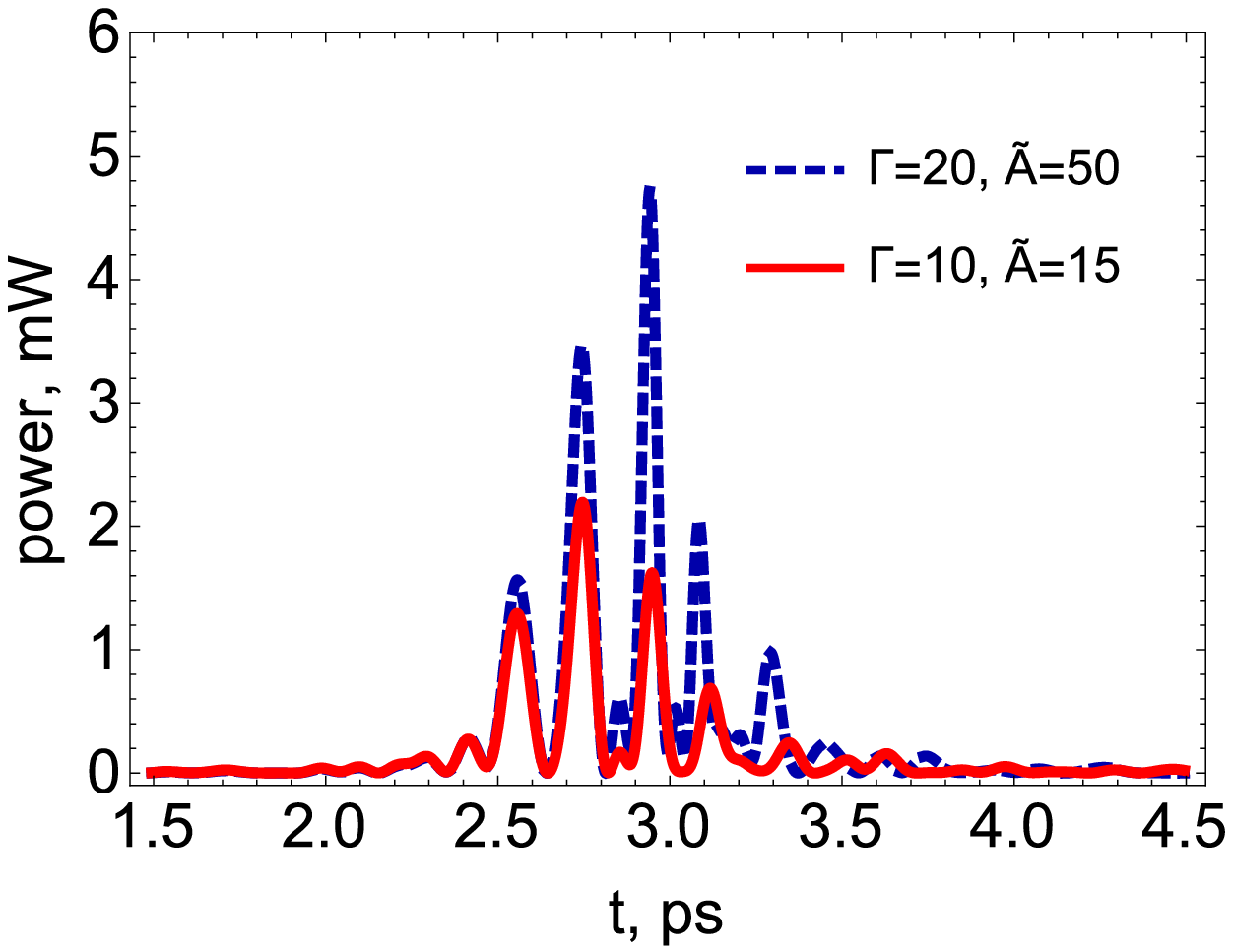}
\caption{\label{Fig_3}
Dependence of signal from the parameters of kernel window transformation function~(\ref{Lorenz}).}
\end{figure}

\begin{figure}[h]
\includegraphics[width=1.65in]{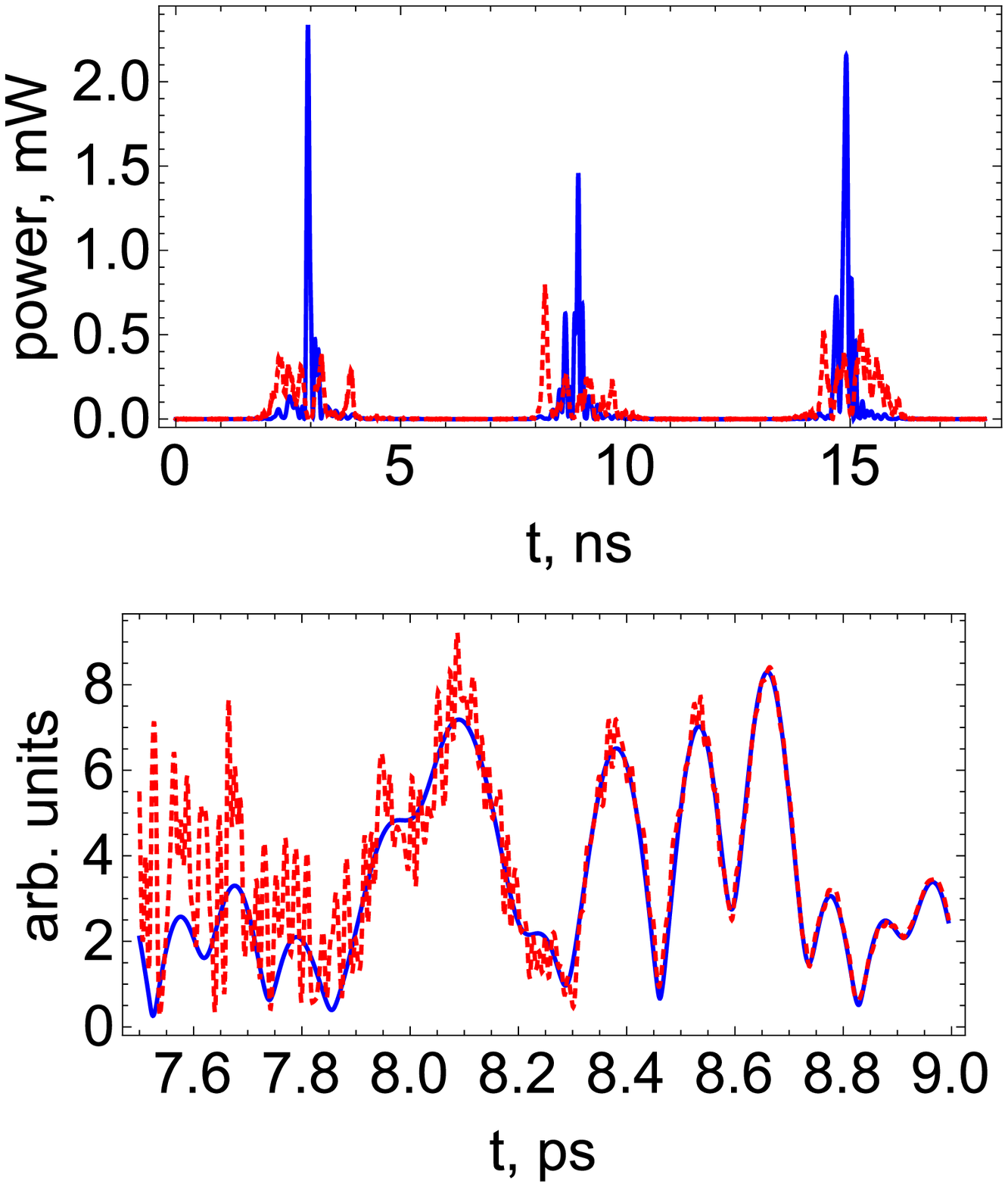}
\includegraphics[width=1.65in]{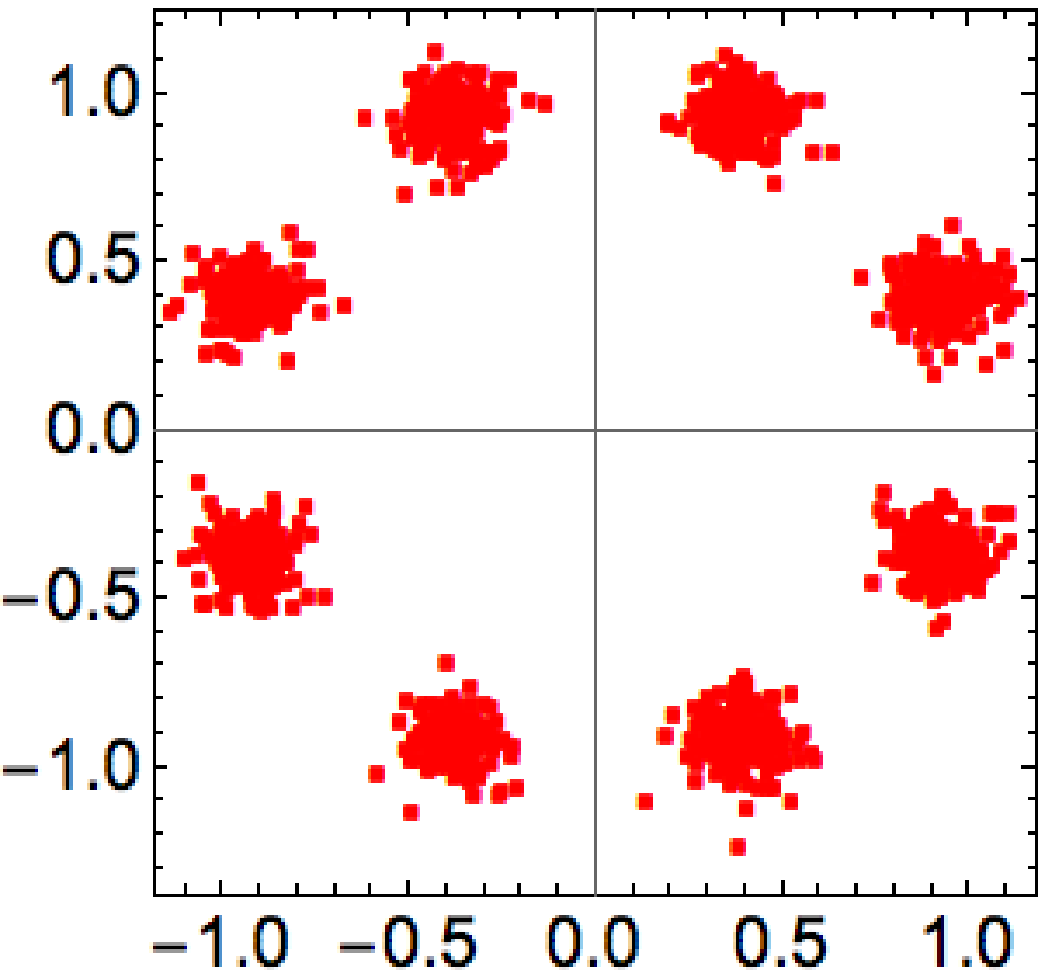}
\caption{\label{Fig_4}
LEFT: Propagation of the burst mode signal in the model NLSE channel with L=1000km and SNR=19.7 dB. Blue solid lines correspond to the signal at the beginning of the line, red dashed lines show the signal at the end of the line. Parameters of of the window transformation function are the same as in the Fig.~\ref{Fig_2}. The bottom picture corresponds to the encoded (blue, solid lines) and decoded using the direct TIB algorithm (red, dashed lines) kernel harmonics for the central burst interval.
\\
RIGHT: Constellation diagram for the central burst interval (statistics on a $10^3$ randomly encoded initial signals).}
\end{figure}

Next, we study the dependence of signal shape on the parameters $\tilde{A}$ and $\Gamma$ of the modulated kernel. We have found that varying the kernel window function parameters allows us to control the characteristics of the generated signal without affecting the information content as illustrated in Fig.~\ref{Fig_3}. Figure~\ref{Fig_4} presents the results of numerical modelling of a burst mode signal transmission in the NLSE channel (with noise) with a total propagation distance L of 1000km and SNR=19.7 dB. The modelling was performed using the standard split-step method with adding noise at each numerical spatial step that corresponds to the distributed noise model (see e.g.~\cite{Rene1,Rene2}). We have found that the direct TIB algorithm remains stable for considered SNR values. Interestingly, better decoding results can be achieved using only the first (left) part of the pulse. This is not surprising since the direct TIB algorithm successively recover the kernel from the left to the right end of the signal. Thus, the right part of the recovered kernel is additionally affected by noise distortions accumulated during the calculation of the left kernel part.

\section{Discussion and Conclusion}

In this work, we proposed and examined new approaches to coding information over the kernel of the GLME. We have considered both the discrete (solitonic) and continuous part of scattering data. We demonstrated that application of the direct TIB method allowes one to recover the most stable part of the kernel, that is an advantage in the presence of distributed noise.

We have proposed, to the best of our knowledge for the first time, to use the general N-soliton solution of the NLSE for simultaneous coding of $N$ symbols
involving $4\times N$ coding parameters, instead of separate $N$ solitons. As a particular sub-class of the general schemes we examined a soliton orthogonal frequency division multiplexing technique that is based on the choice of identical imaginary parts of N-soliton solution eigenvalues, corresponding to equidistant soliton frequencies making it similar to the conventional OFDM scheme. This allows us to use the efficient fast Fourier transform algorithm to recover the data. We would like to point out that efficient implementation of numerical recovery of solitonic kernels by solving GLME requires the development of numerical algorithms, which are stable against additive noise.

For the continuous spectrum we have tested stability of the direct TIB method against the additive noise and proposed to use the localised kernel in the time domain to control properties of the corresponding generated signal. The latter can be considered as a novel realisation of the "nonlinear inverse synthesis" method~\cite{NFT3,le2015nonlinear}.

We demonstrated that the mathematical properties of the NLSE can be used for introducing fundamentally novel (compared to the linear communication theory)
methods for coding and detection of signal setting foundation for the nonlinear communication theory.

\section{Acknowledgments and Contributions}
This work was supported by the UK EPSRC Programme Grant UNLOC EP/J017582/1 and Grant of the Ministry of Education and Science of the Russian Federation (agreement No. 14.B25.31.0003). Part of the work described in the section "Kernel coding of the continuous nonlinear spectrum OFDM" was supported by the Russian Science Foundation (Grant No. 14-22-00174).

All Authors (L.F., A.G., S.K.T.) proposed key ideas, participated in the discussion of results and contributed equally to this work. Numerical modelling was performed by A.G. and L.F.

\bibliographystyle{apsrev4-1}
\bibliography{refs}
\end{document}


\title{The Supplementary Materials for the article: New approaches to coding information using inverse scattering transform}
\date{\today}
\author{L.\,L.~Frumin $^{1,2}$}\email{lfrumin@iae.nsk.su}
\author{A.\,A.~Gelash$^{2,3}$}\email{gelash@srd.nsu.ru}
\author{S.\,K.~Turitsyn$^{2,4}$}\email{s.k.turitsyn@aston.ac.uk}
\affiliation{$^{1}$Institute of Automation and Electrometry SB RAS, Novosibirsk, Russia}
\affiliation{$^{2}$Novosibirsk State University, Novosibirsk, Russia}
\affiliation{$^{3}$Kutateladze Institute of Thermophysics, SB RAS, Novosibirsk, Russia}
\affiliation{$^{4}$Aston Institute of Photonics Technologies, Aston, Birmingham, United Kingdom}
\pacs{02.30.Ik, 05.45.Yv,42.81.Dp,89.70.+c}

\maketitle

\section{Numerical approaches for inverse and direct scattering transform}
In this paragraph we briefly overview and remind basic information concerning numerical Inverse Scattering Transform (IST) which is used in the main text of our paper~\cite{PAPER}. First, we write down the Gelfand-Levitan-Marchenko equations (GLME) in the standard form for the left scattering problem at fixed distance (e.g. $z=0$):
\begin{eqnarray}\nonumber
A^*_1(t,s) + \int_{-s}^t A_2(t,\tau) \Omega(s+\tau) d\tau = 0,
\\\nonumber
-A^*_2(t,s) + \int_{-s}^t A_1(t,\tau) \Omega(s+\tau) d\tau + \Omega(t+s)= 0,
\\\nonumber
\Omega(t) \equiv \Omega(z=0,t),
\\
-t \leqslant s < t, \,\,\,\,\,\,\,\,\,\,\,\,\,\,\, 0 \leqslant t \leqslant T_s.
\label{GLME}
\end{eqnarray}
Here $A_1(t,s)$ and $A_2(t,s)$ are the auxiliary complex functions that links together the kernel $\Omega$ and solution $q$ of the NLSE via the GLME~(\ref{GLME}) and the following relation:
\begin{equation}\label{N-SS}
q(z=0,t) =  -2 A_2^*(t,t) \,.
\end{equation}
The propagation problem is solved by the use of a simple formulae for scattering data evolution (see Eq.~(5) and Eq.~(11) in~\cite{PAPER}).

In the general case numerical solution of an integral equation requires  $\sim M^3$ operations (recall that $M$ is the number of signal discretisation points). To reconstruct the whole signal $q(t_m)$ we need to perform this procedure at all points of the discrete grid (formula.~(7) in~\cite{PAPER}) and, thus, the total cost $\sim M^4$ operations, that is not feasible for practical numerical implementation.

In this work we use the efficient Toeplitz inner-bordering (TIB) numerical scheme for both the inverse and direct scattering transform. Indeed, as it was shown Frumin and co-authors (reference [23] in~\cite{PAPER}) the GLME~(\ref{GLME}) can be rewritten in the Toeplitz form by applying a simple transformation:
\begin{eqnarray}\nonumber
u(t,x) = A_1(t,t-x)\,,
\\
v(t,y) = -A_2^*(t,y-t)\,.
\end{eqnarray}
Now the GLME contains Toeplitz-type kernel $\Omega(y-x)$:
\begin{eqnarray}\nonumber
u(t,x) - \int_{-x}^{2t} \Omega^*(y-x) v(t,y) dy = 0,
\\
v(t,y) + \int_{0}^y \Omega(y-x) u(t,x) dx + \Omega(y)= 0\,,
\label{Toeplitz GLME}
\end{eqnarray}
and, as a result, the numerical TIB IST takes only $M^2$ operations (see details in reference [23] in~\cite{PAPER}). Moreover, recently Frumin and co-authors have demonstrated (reference [24] in~\cite{PAPER}) that the TIB algorithm can be reversely applied to the GLME~(\ref{Toeplitz GLME}), i.e. it allows to find the kernel  $\Omega(t_m)$ from the known signal $q(t_m)$. Again, the required number of numerical operations is $M^2$.  The numerical schemes and details can be found in references [23,24] in~\cite{PAPER}.

Here~\cite{PAPER} we apply both inverse and direct TIB algorithm to the continuous spectrum signals. For the discrete spectrum case we apply only direct TIB method to recover the kernel, meanwhile to create signal at the beginning of the transmission line we use exact N-SS, described in the next paragraph.
\subsection{$N$-soliton solutions of the NLSE}

For the discrete spectrum kernel (see formula (3) in~\cite{PAPER}) factorization of the GLME~(\ref{GLME}) leads to the system of linear algebraic equations (see, for instance the monograph of Lamb~\cite{lamb1980elements}). Then, the N-SS can be found in the following exact form:

\begin{equation}\label{N-SS}
q^{(N)}(z=0,t) = -2 \bra{\mathbf{\Psi}(t)}(\widehat{\mathbf{E}} +
\widehat{\mathbf{M}}(t)^*
\widehat{\mathbf{M}}(t))^{-1}\ket{\mathbf{\Phi}(t)}\,.
\end{equation}
Here $\widehat{\mathbf{E}}$ is $N\times N$ identity matrix,
\begin{eqnarray}\label{N-SS details}
\bra{\mathbf{\Psi}(t)} = \bra{c_1 e^{-i\xi_1
t},...,c_Ne^{-i\xi_1 t}}\,,
\\\nonumber
\bra{\mathbf{\Phi}(t)} = \bra{e^{-i\xi_1 t},...,e^{-i\xi_1
t}}\,,
\\\nonumber
\widehat{\mathbf{M}}_{k,j}(t) = c_j \frac{e^{i (\xi^*_k-\xi_j)
t}}{\xi^*_k-\xi_j}\,,
\end{eqnarray}
and parameters $c_k$ are defined in~\cite{PAPER}.

To the best of our knowledge all the existing discrete spectrum numerical IST algorithms are unstable at large $N$, that can be understood by looking at the exact N-SS formulae~(\ref{N-SS}),(\ref{N-SS details}). Indeed, the eigenvalues $\xi_k$ are complex and, thus, the matrix $\widehat{\mathbf{E}} +\widehat{\mathbf{M}}(t)^*\widehat{\mathbf{M}}(t)$ in~(\ref{N-SS}) may become ill-conditioned at large $|t|$. In such cases we use the arbitrary precision arithmetics to obtain accurate N-SS signal. Recently, A.A.Gelash and D.S. Agafontsev found that numerical realisation of the Zakharov-Shabat dressing method can be stably used up to $N \sim 32$ soliton solutions~\cite{gelash2016uniform}. Application of the dressing method to our kernel-based approach is a nontrivial task, however we believe that this can be an interesting direction for future research.

 \section{Parametric kernel decoding}
In this paragraph we discuss the N-SS kernel general parametric encoding/decoding schemes involving $4\times N$ coding parameters.
Let us write the N-SS kernel (formula~(3) in~\cite{PAPER}) as a time series on the discrete grid (see formula.~(7) in~\cite{PAPER}):
\begin{eqnarray}\label{kernel sampling}
\Omega_m \equiv \Omega(t_m) =
\\\nonumber
=\sum_{k=1}^{N} c_k e^{-i\xi_k t_m} = \sum_{k=1}^{N} c_k e^{-i\xi_k T(m-1)} = \sum_{k=1}^{N} c_k z_k^{m-1}\,.
\end{eqnarray}
Parameters $z_k=\exp(-i\xi_k T)$ in~(\ref{kernel sampling}) are defined by the soliton eigenvalues $\xi_k$ and by the value of time slot $T$. Here, we again choose the minimum possible number of time samples $M=N$. Then, for the decoding problem we obtain system of equations with the Vandermonde matrix:
\begin{equation}\label{vandermonde}
\left(\begin{array}{cccc}z_1^0, & z_2^0, & ..., & z_N^0 \\z_1^1, & z_2^1, & ..., & z_N^1 \\... & ... & ... & ... \\z_1^{N-1}, & z_2^{N-1}, & ..., & z_N^{N-1},\end{array}\right)
\left(\begin{array}{c}c_1 \\c_2 \\... \\c_N\end{array}\right)=
\left(\begin{array}{c}\Omega_1 \\\Omega_2 \\... \\\Omega_N\end{array}\right)\,.
\end{equation}
Now we consider both \textit{position-phase} modulation and \textit{amplitude-frequency} modulation of the N-SS kernel and discuss numerical problems that occur in general case.

\subsection{Position-phase modulation}
Suppose we know the eigenvalues $\xi_k$ and hence the parameters
$z_k = \exp(-i\xi_k)$. The decoding problem is to find the
parameters $c_k$ by the measured kernel samples $\Omega_i$, that
can be done by solving system~(\ref{vandermonde}). However, the
straightforward numerical algorithm based, for example, on Gauss
elimination in a general case is extremely challenging since the Vandermonde matrix~(\ref{vandermonde})
exponentially fast becomes ill-conditioned with the increase of
$N$~\cite{tyrtyshnikov1994bad}. On the other hand, the Vandermonde
matrix belongs to the class of structured matrices for which the
effective numerical algorithms have been
developed~\cite{blahut2010fast}. By applying the effective
matrix inversion algorithm the kernel decoding can be
performed using $N^2$ operations, however, the numerical stability
restricts $N$  by around $\sim 60$ harmonics (see, for example~\cite{gohberg1997fast}).

The inversion of the Vandermonde matrix becomes numerically stable at any $N$ when $z_k$ are the
complex $k$-th roots of unity. For the N-SS kernel this is possible only when $\xi_k$ have
identical imaginary parts (that can be moved to the right part of the matrix system~(\ref{vandermonde})), i.e. in the case presented by formula (6) in~\cite{PAPER}.
The additional harmonics orthogonality condition (formula.~(8) in~\cite{PAPER}) allows us to use the FFT/IFFT algorithms instead of matrix inversion operations, that motivated us to focus on this elegant encoding scheme~\cite{PAPER}.

\subsection{Amplitude-frequency modulation}
Another possibility is to use the eigenvalues $\xi_k$ as the carriers of
information. They have to be found from the measured kernel samples
$\Omega_m$, while the shift-phase parameters $c_k$ are all known and are not used for coding of information. The
parametric approach based on the Prony's method (see, for
example~\cite{marple1987digital}, chapter 11) uses the following
\textit{master polynomial}
\begin{equation}\label{polynomial}
\phi(z) = \prod_{n=1}^N (z-z_k)^n = \sum_{n=0}^N a_n z^n, \,\,\,\,\,\, a_0=1\,,
\end{equation}
with the complex roots $z_k$. The coefficients $a_n$ of the
polynomial~(\ref{polynomial}) can be determined by solving the Toeplitz
system of equations:
\begin{equation}\label{toeplitz}
\left(\begin{array}{cccc}\Omega_N, & \Omega_{N-1}, & ..., & \Omega_1 \\ \Omega_{N+1}, & \Omega_N, & ..., & \Omega_2 \\... & ... & ... & ... \\ \Omega_{2N-1}, & \Omega_{2N-2}, & ..., & \Omega_N,\end{array}\right)
\left(\begin{array}{c}a_1 \\a_2 \\... \\a_N\end{array}\right)=
\left(\begin{array}{c}\Omega_{N+1} \\\Omega_{N+2} \\... \\\Omega_{2N}\end{array}\right)\,.
\end{equation}
Numerical solution of the problem~(\ref{toeplitz}) can be obtained
by the use of Levinson-Durbin-Trench algorithm through the
$O(N^2)$ arithmetic operations~\cite{blahut2010fast}. However, the
subsequent roots finding of the master polynomial $\phi(z)$ is the
hard numerical problem for the large number of samples $N$. For
example, the well known factorization algorithm of Jenkins and
Traub becomes numerically unstable at $N \sim 100$
\cite{jenkins1972algorithm}.

We note, that in the case of continuous spectrum
kernel (formula (10) in~\cite{PAPER}) the corresponding Vandermonde matrix
can be always stably inverted since it becomes Fourier matrix.

We conclude that the general (parametric) N-SS kernel decoding requires matrix inversion and/or finding roots of the polynomial in the decoder. Although, the advanced numerical algorithms with a relatively small number of operations $\sim N^2$ can be exploited, their stability against large number of harmonics and additive noise requires a separate comprehensive analysis that is beyond the scope of this Letter.

\bibliographystyle{apsrev4-1}
\bibliography{refs}